\documentclass[useAMS,usenatbib]{mn2e}
\usepackage{epsfig}
\usepackage{graphicx}
\usepackage{epstopdf}

\def\simgt{\lower.5ex\hbox{$\; \buildrel > \over \sim \;$}}
\def\simlt{\lower.5ex\hbox{$\; \buildrel < \over \sim \;$}}
\def\sp{{\it SP }}
\def\rtwopf{{\it r2p5}}
\def\rfive{{\it r5}}
\def\rten{{\it r10}}
\def\rtwentyf{{\it r25}}

\newcommand\aj{{AJ}}%
\newcommand\apj{{ApJ}}%
\newcommand\aap{{A\&A}}%
\newcommand\mnras{{MNRAS}}%
\title[Binary star disruption]{Binary Star Disruption in Globular
  Clusters with Multiple Stellar Populations}
\author[E. Vesperini et al.]  {Enrico Vesperini$^1$\thanks{E-mail:
    vesperin@physics.drexel.edu}, Stephen L.W. McMillan$^1$, Francesca
  D'Antona,$^2$
  Annibale D'Ercole,$^3$\\
  $^1$Department of Physics, Drexel University, Philadelphia, PA 19104, USA\\
  $^{2}$INAF- Osservatorio Astronomico di Roma, via di Frascati 33,
  I-00040 Monteporzio (Italy)\\
  $^{3}$INAF- Osservatorio Astronomico di Bologna, via Ranzani 1,
  I-40127 BOLOGNA
  (Italy)\\
}
\begin{document}
\date{Accepted ... Received ...; in original form ...}
\maketitle

\label{firstpage}

\begin{abstract}
  The discovery of multiple stellar populations in globular clusters
  raises fundamental questions concerning the formation and dynamical
  history of these systems.  In a previous study aimed at exploring
  the formation of second-generation (SG) stars from the ejecta of
  first-generation (FG) AGB stars, and the subsequent dynamical
  evolution of the cluster, we showed that SG stars are expected to
  form in a dense subsystem concentrated in the inner regions of the
  FG cluster.  In this paper we explore the implications of the
  structural properties of multiple-population clusters, and in
  particular the presence of the inner SG subsystem, for the
  disruption of binary stars.  We quantify the enhancement of the
  binary disruption rate due to the presence of the central SG
  subsystem for a number of different initial conditions.  Our
  calculations show that SG binaries, which are assumed to be more
  concentrated in the cluster inner regions, are disrupted at a
  substantially larger rate than FG binaries.  Assuming a similar
  initial fraction of FG and SG binaries, our dynamical study
  indicates that the SG population is now expected to contain a
  significantly smaller binary fraction than the FG population.

\end{abstract}


\section{Introduction}
\label{sec:intro}
Spectroscopic and photometric observations reveal that globular
clusters host multiple stellar populations.  Spectroscopic studies
show star-to-star variations in the abundances of light elements, such
as Na, O, Al, and Mg, indicating that a significant fraction (50-80\%)
of globular cluster stars must have formed out of matter processed
through a high-temperature CNO cycle in a first generation (hereafter
FG) of stars (see e.g. Carretta et al. 2009a, 2009b and references
therein).  Photometric studies reveal the presence of multiple main
sequences, subgiant, and red-giant branches in numerous clusters,
supporting the spectroscopic evidence for multiple populations (see
e.g. Piotto 2009 and references therein).  Photometric measurements
also provide evidence for a population of very He-rich stars among the
second-generation (hereafter SG) stars of some clusters (see
e.g. Piotto et al. 2007).  These findings confirm the predictions of
previous studies (D'Antona et al. 2002; D'Antona \& Caloi 2004,
D'Antona \& Caloi 2008) that suggested the existence of a population
of stars with a strong He enhancement on the basis of the extension of
the horizontal branch.

The origin of the gas out of which SG stars formed is still an open
question: in addition to the scenario involving AGB stars (Cottrell \&
Da Costa 1981, Ventura et al. 2001) described in more detail below,
possible gas sources proposed in the literature also include rapidly
rotating massive stars and massive binary stars (Decressin et
al. 2007, De Mink et al. 2009; see also Renzini 2008 and references
therein for a review).  Many fundamental questions concerning globular
cluster star formation and cluster chemical and dynamical history are
raised by the discovery of multiple populations, and are targets of
ongoing investigations (see e.g. D'Ercole et al. 2008, 2010, Vesperini
et al. 2010, Bekki 2011 and references therein).

D'Ercole et al. (2008) explored the formation and dynamical evolution
of multiple populations in globular cluster by means of hydrodynamical
and N-body simulations, focusing on a model in which SG stars form
from the ejecta of FG AGB stars.  Our simulations show that the AGB
ejecta form a cooling flow and rapidly collect in the innermost
regions of the cluster, forming a concentrated SG stellar subsystem.
In order to form the numbers of SG stars observed today, the FG
cluster must have been considerably more massive than it is now, and
the majority of stars in the cluster initially belonged to the FG
population.  The N-body simulations presented in D'Ercole et
al. (2008) show that the early expansion triggered by the loss of mass
in the form of SNII ejecta and primordial gas leads to a strong
preferential loss of FG stars.

According to our models, during this early evolutionary phase the
cluster evolves from a configuration in which FG stars dominate to one
in which the numbers of FG and SG stars are similar (or even one in
which the SG stars are now the dominant population), as observed in
several Galactic globular clusters (see e.g. Carretta et al. 2009a,
2009b).  At the end of this early phase, the SG subsystem, although
now composed of a number of stars similar to that of the FG
population, is still concentrated in the cluster inner regions.  The
system thus begins its long-term, relaxation-driven evolution with a
structure characterized by the superposition of a compact SG inner
cluster and a more extended FG cluster.  In the initial simulations
discussed in D'Ercole et al. (2008) we showed that, as the cluster
evolves, the two populations tend to mix and the characteristic
structure imprinted by the SG formation process is slowly
erased. 

Such a peculiar structure differs significantly from the simple King
or Plummer models usually adopted as initial conditions in studies of
globular cluster evolution.  It is therefore important to explore the
implications of this unusual initial state for the long-term evolution
of a cluster's structural and kinematical properties, as well as its
stellar content.  This paper focuses on the effects of this new class
of initial structural properties on the survival of binary stars, and
on the differences in the evolution of the numbers of FG and SG
binaries.

Binary stars play a crucial role in cluster evolution, both as an
energy source supporting the cluster's post-core collapse dynamics
(see e.g. Gao et al. 1991, Goodman \& Hut 1989, Vesperini \& Chernoff
1994, Heggie et al. 2006, Trenti et al. 2007, Hurley et al. 2007; see also Heggie \& Hut
2003 for a review), and as potential seeds for the formation of a
variety of exotic objects (e.g. LMXBs, CVs, blue stragglers etc.; see
e.g. Ivanova et al. 2006, 2008, Ferraro \& Lanzoni 2008 and references
therein).  Exploring the evolution of binaries in the environment
resulting from the formation of SG stars, and understanding whether or
not observed binary properties and abundances contain any imprint of a
cluster's formation, are fundamental problems to be addressed.  Here
we present an initial study of this problem that combines the results
of a set of N-body simulations of clusters hosting multiple
populations with analytical expressions for binary interaction rates,
to estimate the net binary disruption rate during cluster evolution.

The structure of this paper is as follows: in Section 2 we describe
the N-body simulations and the analytical framework used for our
study; in Section 3 we present our results; and in Section 4 we
discuss our results and summarize our main conclusions.

\section{Analytical and numerical framework}
\label{sec:framework}
\subsection{Binary star disruption}
\label{sec:binardisr}
The outcome of the interactions between a binary and single stars in a
cluster depends critically on the ratio $x=\epsilon/m\sigma^2$ of the
binary binding energy, $\epsilon=Gm^2/2a$ to the mean kinetic energy
of cluster stars, $m \sigma^2$, where $m$ is the mass of a single star
and each of the binary components (assumed here to be the same), $a$
is the binary semi-major axis, and $\sigma$ is the 1-D stellar
velocity dispersion (notice that $m \sigma^2$ is often also referred
to as $kT$ in the literature).  Substantial analytical and numerical
effort has been devoted to the study of binary-single star
interactions and their possible outcomes (see e.g. Heggie \& Hut 2003
for a review and references therein).  Binaries having $x \simgt 1$
(so called {\it hard} binaries) will, on average, become more bound as
a result of interactions with single stars; for $x \simlt 1$ ({\it
  soft} binaries), binary--single-star interactions will generally
lead to the disruption of the binary star.

The goal of this paper is to explore the extent to which the presence
of a compact SG subcluster in the inner regions of a young globular
cluster can increase the disruption rate of binaries.  We focus here
on the process of binary {\it ionization}, that is on the disruption
of a binary following a single interaction with a single star. The
disruption rate of a population of binaries with number density $n_b$,
semi-major axis $a$ interacting with a population of single stars with
number density $n_s$ can be written
\begin{equation}
\label{eq:ion}
    {1\over n_b}{dn_b\over dt}=n_s \pi a ^2 V_{th} R(x),
\end{equation}
where $R(x)$ is an analytical fit obtained by Hut \& Bahcall (1983)
from a series of numerical simulations of binary-single star
encounters:
\begin{equation} 
    R(x)={1.64 \over (1+0.2A/x)[1+\exp(x/A)]},
\end{equation}
and $V_{th}=3(A/2)^{1/2}\sigma$ is the dispersion of
binary--single-star relative velocities.  The quantity $A$ is equal to
1 in the case of energy equipartition between the binary and the
single star populations, and equal to $4/3$ in the case of velocity
equipartition.  As a cluster evolves toward energy equipartition, the
appropriate value of $A$ will therefore change from $4/3$ to $1$.
Hereafter, we adopt $A=1$ as our default value.

In this study we follow the evolution of the overall binary disruption
rate, and the differences in the disruption rates for SG and FG
binaries, by combining Eq.\ref{eq:ion} with the time-evolving radial
profiles of $n_s$ and $\sigma$ obtained from our N-body simulations of
multiple-population clusters.

In our calculations we assume that the spatial distribution of SG (FG)
binaries always follows the distribution of SG (FG) single stars,
$n_{b,SG}=(N_{b,SG}/N_{s,SG})n_{s,SG}$ (and similarly for FG).  We
note that, by making this assumption, we explicitly neglect any
possible spatial segregation of the binary population and the ensuing
increase in binary disruption.  This neglect will be most significant
for the initially most compact SG population, and will further
increase the preferential disruption of SG binaries described below.
Other processes whose effects are not included here are binary-binary
interactions and the possible formation of binaries by three-body and
two-body tidal encounters.  Finally, for clusters with a significant
fraction of primordial binaries, binary heating will delay deep core
collapse and the binary disruption occurring during that high-density
phase.  (Notice, however, that a larger fraction of binaries delaying
deep core collapse will increase the number of binary-binary
interactions, again leading to significant binary disruption).

In order to quantify the binary disruption rate, we define the radial
profile of the ionization rate
\begin{equation}
    I(r,a)=n_s(r) \pi a^2 V_{th}(r) R(x)
\label{eq:ionrate}
\end{equation}
and the volume-integrated disruption rate for SG (FG) binaries
\begin{eqnarray}\nonumber
\label{eq:intdis}
    {1\over N_{b,SG}}{\hbox{d} N_{b,SG} \over \hbox{d} t} &\equiv& \Delta_{SG}(a,t) \\
	&=& {1\over N_{s,SG}}\int n_{s,SG}(r)n_s(r) \pi a ^2 V_{th}(r) R(x) \nonumber\\
	&& ~~~~~~~~~~\times 4 \pi r^2 \hbox{d} r;
\end{eqnarray}
(and similarly for $\Delta_{FG}$), where $a$, $x$, and $ V_{th}$ are
defined above.

Finally, we calculate the time integral of Eq.\ref{eq:intdis} to
obtain the time evolution of the number of SG binaries with a given
semi-major axis $a$ as 
\begin{equation}
   { N_{b,SG}(a,t)\over N_{b,SG}(a,0)}
	= exp\left[-\int_0^t \Delta_{SG}(a,\tilde{t}) \hbox{d} \tilde{t}\right]
\end{equation}
and similarly for $N_{b,SG}(a,t)$.

\subsection{N-body simulations}
\label{sec:nbody}
As discussed in Section \ref{sec:intro}, our previous hydrodynamical
and N-body simulations (D'Ercole et al. 2008) predict that SG stars
form in a compact subsystem concentrated in the inner cluster regions
(see also Bekki 2011 for additional simulations confirming our
prediction), and that early cluster expansion triggered by the loss of
SNII ejecta and primordial gas is responsible for the loss of a
significant fraction of FG stars.

In order to explore the long-term evolution of a multiple-population
cluster, driven by two-body relaxation, we have followed the evolution
of four different systems each containing equal numbers of SG and FG
stars.  In all cases, the initial state of the FG system is modeled as
a single-mass King (1966) model with central dimensionless potential
$W_0=7$ and truncation radius $R_t$ equal to the cluster Jacobi
radius.  The initial SG subsystem is also modeled as a $W_0=7$ King
model, but one that is concentrated in the inner regions of the FG
cluster.  We have explored the evolution of the resulting
two-population cluster for four values of the initial ratio of the FG
to SG half-mass radii: $R_{h,FG}/R_{h,SG}=2.5,~5,~10,~25$.  We refer
to these simulations as \rtwopf, \rfive, \rten, and \rtwentyf,
respectively.  The different ratios represent clusters with the same
initial mass and total radius ($R_t$) but different internal
structures, as measured by the degree of concentration of the SG
subsystem.

All simulations start with a total number of particles $N=10,000$, and
our analysis focuses on the evolution of the system until $t\sim 40
t_{rh,SG}(0)$, where $t_{rh,SG}(0)$ is the initial half-mass
relaxation time of the SG subsystem.  For all initial conditions
considered, the system undergoes core collapse at $t\approx 17-25
t_{rh,SG}(0)$.  More detailed discussion of the structural evolution
of the FG and SG systems will be presented in a separate paper
(Vesperini et al. 2011, in preparation).

\section{Results}
We consider a common range of binary semi-major axes, $10^{-2}
R_t/N<a<10^{-1} R_t/N$, for all simulations.  To illustrate the effect
of the inner SG substructure in our multiple-population clusters, in
some cases we compare quantities related to binary disruption with
those calculated for a 'standard' $W_0=7$ King model with no multiple
populations and hence no additional substructure due to the presence
of the SG system (hereafter we will refer to this system as the
\sp-Single Population- system).  For the \sp system, $R_h\sim 0.1R_t$
and the above range of $a$ corresponds to binaries that are close to
the hard/soft boundary or are hard everywhere in the cluster;
specifically for the widest binary considered, $x \simgt 2.8$ in the
\sp cluster.

\begin{figure}    
\centering{
\includegraphics[width=8cm]{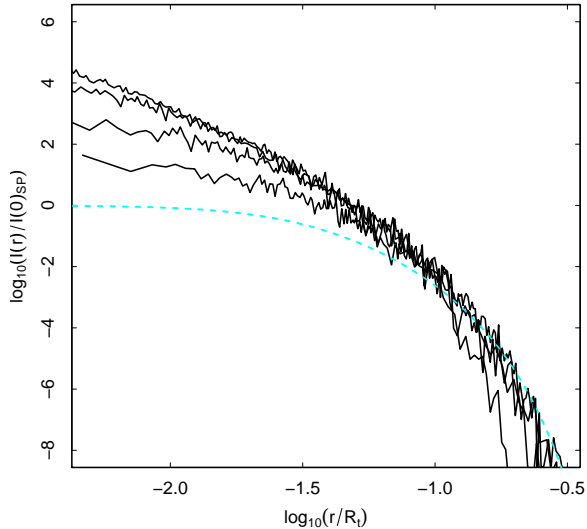}
}
\caption{Radial profile of the ionization rate, $I(r)$ (see
  Eq.\ref{eq:ionrate}) calculated for a binary with semi-major axis
  $a=0.05R_t/N$ for the four systems considered, normalized to the
  central value of the \sp system.  The cyan dashed line shows the
  profile for the \sp system; the solid lines show, from top to
  bottom, the profiles for the \rtwentyf, \rten, \rfive, and \rtwopf~
  systems.}
\label{fig:ionrate} 
\end{figure}

Fig.\ref{fig:ionrate} shows the radial profile of the ionization rate
$I(r)$ for $a=5\times 10^{-2}R_t/N$ in each of the models explored.
The profiles are calculated using the cluster properties after about
5-6 dynamical times, in order to allow the initial
two-population configuration to come into dynamical equilibrium.  This
figure illustrates how the presence of the inner SG system greatly
enhances the ionization rate over that in the simple \sp system.  Even
for the system (\rtwopf) with the least concentrated SG population,
the ionization rate in the cluster inner regions is about an order of
magnitude larger than that in the \sp system.

\begin{figure}    
\centering{
\includegraphics[width=8cm]{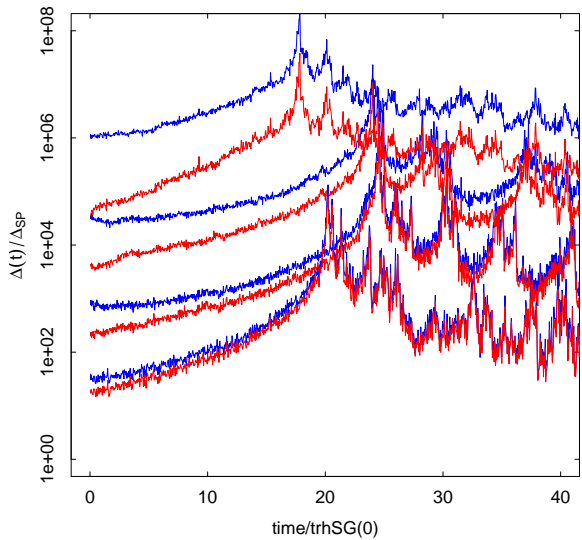}
}
\caption{Time evolution of the volume-integrated ionization rate
  (normalized to the value for the \sp system) for SG,
  $\Delta_{SG}(a,t)$ (blue lines), and FG binaries, $\Delta_{FG}(a,t)$
  (red lines), for $a=5\times 10^{-2} R_t/N$.  Each set (red and blue)
  of lines refers, from top to bottom, to the \rtwentyf, \rten,
  \rfive, and \rtwopf~ systems.}
\label{fig:volumeionrate} 
\end{figure}

\begin{figure}    
\centering{
\includegraphics[width=8cm]{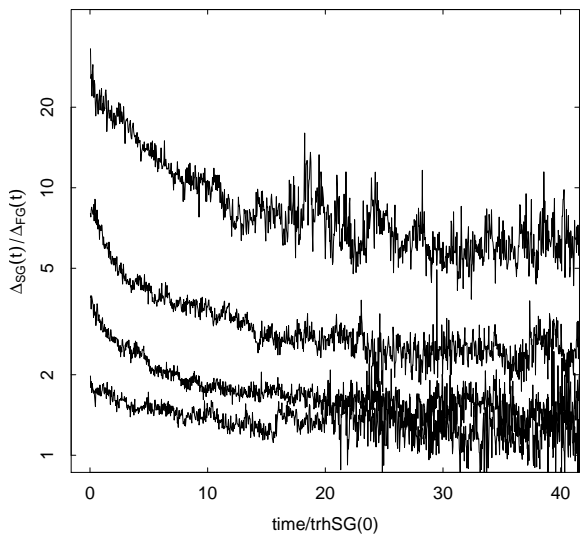}
}
\caption{Time evolution of the ratio of the SG to the FG
  volume-integrated ionization rate
  $\Delta_{SG}(a,t)/\Delta_{FG}(a,t)$, for $a=5\times 10^{-2} R_t/N$.
  From top to bottom, the four lines refer to the \rtwentyf, \rten,
  \rfive, and \rtwopf~ systems.}
\label{fig:ratiorate} 
\end{figure}

As the system evolves, the ionization rate reaches a maximum at the
time of core collapse and decreases again in the post-core-collapse
phase.  However, the ionization rate in all multiple-component models
is always larger than that in the reference \sp system.
Fig.\ref{fig:volumeionrate} shows the time evolution of the
volume-integrated ionization rate for SG ($\Delta_{SG}(a,t)$) and FG
($\Delta_{FG}(a,t)$) binaries, for $a=5\times 10^{-2} R_t/N$.  SG
binaries are preferentially located in the cluster inner regions and
so are disrupted more efficiently than FG binaries.  The time
evolution of $\Delta_{SG}(a,t)/\Delta_{FG}(a,t)$ is plotted in
Fig.\ref{fig:ratiorate}.

As the cluster evolves and the two populations tend to mix,
$\Delta_{SG}(a,t)/\Delta_{FG}(a,t)$ tends to decrease.  However, the
mixing is not complete by the end of the time interval spanned by this
study ($t\sim 40 t_{rh,SG}(0)$), so the SG binary disruption rate is
always larger than that of FG binaries.

\begin{figure}    
\centering{
\includegraphics[width=8cm]{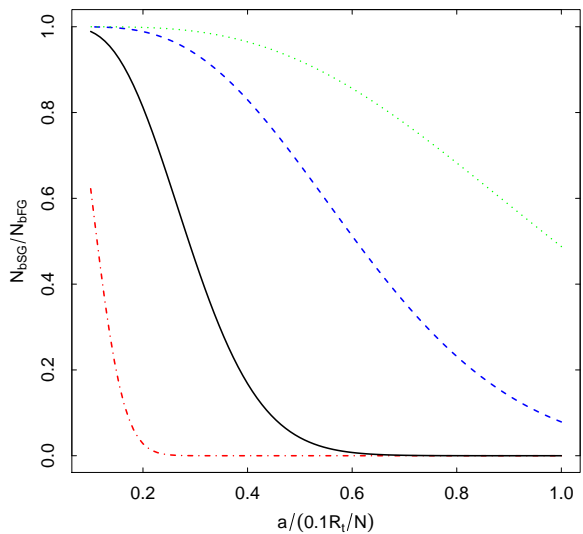}
}
\caption{Ratio of the number of SG to FG binaries, at $t=40
  t_{rh,SG}(0)$, versus binary semi-major axis, $a$, for the
  \rtwentyf~ (red dot-dashed line), \rten~ (black solid line), \rfive~
  (blue dashed line), and \rtwopf~ (dotted green line). }
\label{fig:numberratio} 
\end{figure}

\begin{figure}    
\centering{
\includegraphics[width=8cm]{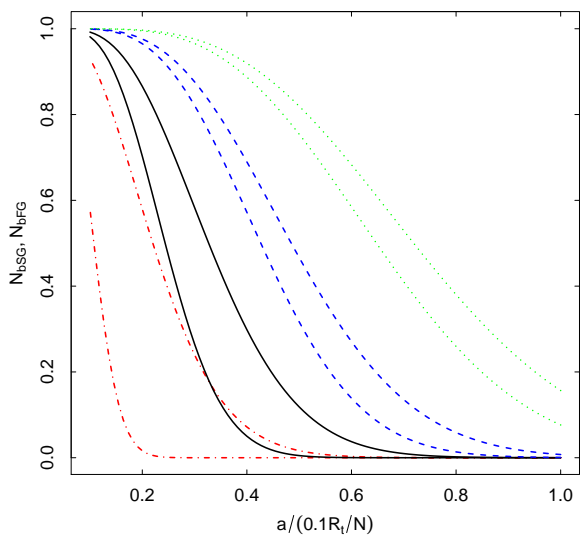}
}
\caption{Numbers of SG and FG binaries (normalized to their initial
  values) at the end of the simulations, versus binary semi-major
  axis, $a$, for the \rtwentyf~ (red dot-dashed line), \rten~ (black
  solid line), \rfive~ (blue dashed line), and \rtwopf~ (dotted green
  line) runs.  For each set of lines, the upper (lower) curve refers
  to the FG (SG) binary population.}
\label{fig:numberbinaries} 
\end{figure}

Fig. \ref{fig:numberratio} plots $N_{b,SG}(a)/N_{b,FG}(a)$, calculated
at  the end of each simulation, as a function of binary semi-major
axis.  In all cases, SG binary disruption is significantly enhanced
compared to that of the FG binary population.  Fig.
\ref{fig:numberbinaries} shows the ratios of the final to the initial
number of binaries for the SG and the FG population,
$N_{b,SG}(a)/N_{b,SG}(a)_{init}$ and $N_{b,FG}(a)/N_{b,FG}(a)_{init}$,
and illustrates the extent of the binary disruption in our multiple
population clusters, as well as the preferential disruption of SG
binaries.

\begin{figure}    
\centering{
\includegraphics[width=8cm]{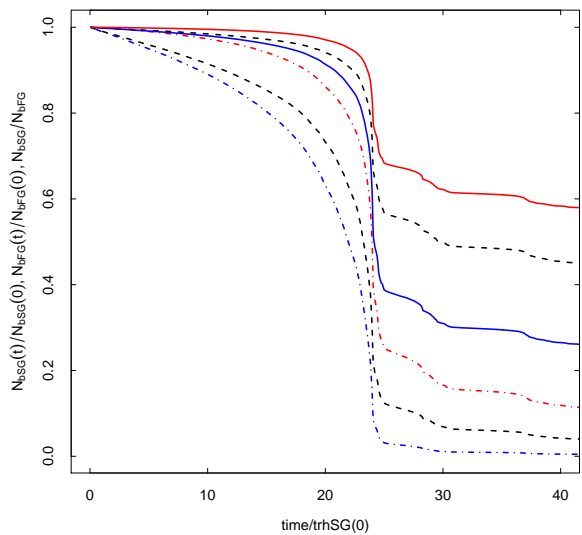}
}
\caption{Time evolution of the numbers of SG and FG binaries
  (normalized to their initial value) and their ratio, for simulation
  \rten~ for FG binaries and $a=3\times 10^{-2} R_t/N$ (upper red solid
  line), SG binaries and $a=3\times 10^{-2} R_t/N$ (lower blue solid
  line), FG binaries and $a=5\times 10^{-2} R_t/N$ (upper red
  dot-dashed line), SG binaries and $a=5\times 10^{-2} R_t/N$ (lower
  blue dot-dashed line).  The upper black dashed line shows the ratio
  of the number of SG to FG binaries for $a=3\times 10^{-2} R_t/N$;
  the lower black dashed line shows the SG to FG binary number ratio
  for $a=5\times 10^{-2} R_t/N$. }
\label{fig:timenumber} 
\end{figure}

Finally, in Fig.\ref{fig:timenumber} we plot the time evolution of
$N_{b,SG}(a,t)$, $N_{b,FG}(a,t)$, and their ratio, for two different
values of $a$ for simulation \rten.  This figure illustrates the
extent of the early (pre-core collapse) disruption of binaries due to
the presence of the high-density SG subsystem.  Binary disruption
further increases during core collapse, and finally slows down during
the post-core collapse phase.  The preferential disruption of SG
binaries continues for the whole simulation.

\section{Discussion and Conclusions}
The results presented in this paper show that the properties of binary
stars are significantly affected by the initial structure of a cluster
hosting multiple populations, and may contain important clues to the
formation and evolutionary history of multiple populations.

The central result of our study is that significant differences in the
numbers of FG and SG binaries is a fingerprint of the structural
properties predicted by our models of SG star formation and dynamical
evolution (D'Ercole et al. 2008; further investigation of the
long-term evolution of multiple population clusters will be presented
in Vesperini et al. 2011, in preparation).  Specifically, we showed in
our previous studies that SG stars forming from AGB ejecta tend to
form in a strongly concentrated subsystem in the FG cluster inner
regions (see also Bekki 2011); we have now shown in this paper that in
a cluster with such a structure, SG binaries are preferentially
disrupted and that, more generally, binary disruption is enhanced
compared to a standard cluster with similar mass and size but without
an inner SG subsystem.

Our calculation is based on analytical calculations combined with the
results of N-body simulations of cluster structural and kinematical
evolution.  Further refinement of the calculations presented in this
paper will require simulations including full treatment of binary
stars; simulations including the self-consistent treatment of binaries
are very computationally expensive and are beyond the scope of this
initial exploratory study.  Future studies based on simulations with
binaries will allow us to incorporate additional effects
(e.g. binary-binary interactions, binary segregation, binary heating)
not included in our simple analytical treatment of binary disruption.

The first observational indication of the preferential disruption of
SG binaries, as found in our analysis, comes from a recent study of Ba
stars in globular clusters (D'Orazi et al. 2010).  Ba stars are
thought to be the result of the accretion of matter processed by a
thermally pulsing AGB onto the secondary component of a binary
system. The Ba-rich envelope of the AGB component contaminates the
companion via wind accretion, possibly followed by stable Roche lobe
overflow, or by common envelope evolution, depending on the mass ratio
at the time the donor AGB fills its Roche lobe (see Han et al. 1995
for a detailed study of the binary channels for the formation of Ba
stars; see also Jorissen et al. 1998, McClure \& Woodsworth 1990 for
two observational studies of the Ba stars orbital properties).

In order for this process to occur, the initial binary separation must
be larger than approximately 1 AU (smaller separations would affect
the primary component evolution before it reaches the AGB phase).
D'Orazi et al. find that Ba stars belong predominantly to the FG
population.  This result is consistent with a dynamical history in
which the SG Ba star binary progenitors are disrupted more efficiently
than those of the FG population.

We note that the larger binary interaction rate of SG binaries can
also lead them to harden more rapidly.  By tightening a binary below
the minimum separation for the formation of Ba stars, the more rapid
hardening of SG binaries may represent an additional channel for the
suppression of Ba stars in the SG population.  Another interesting
possible consequence of an enhanced binary interaction rate is that a
binary might be disrupted after the mass transfer episode that led to
the formation of the Ba star, resulting in a single Ba star.

The sample of Ba stars discussed in D'Orazi et al. is still very
small, and further observational studies aimed at exploring the
relative abundance of FG and SG binaries will be extremely important
to test the predictions of our study and, more generally, to shed
further light on the interplay between the dynamics of multiple
population clusters and their binary star content.

Many observational studies cover only a limited region of a cluster,
and thus could provide only local information on the SG-to-FG binary number
ratio.  A detailed comparison between observations and
theoretical results will entail a more detailed accounting of the
radial variation of the SG and FG binary fractions.  The results
presented here illustrate a clear general trend in the evolution of
the total numbers of SG and FG binaries, but the interplay between
binary segregation and binary disruption, along with possible initial
differences in the spatial distributions of FG and SG binaries, will
presumably result in radial gradients in the SG-to-FG binary number
ratio.  (For example, in the innermost regions, the binary population
might be dominated by SG binaries even though FG binaries may dominate
globally.)  As discussed above, future simulations including
self-consistent treatments of binaries will include these effects,
allowing us to further explore these issues.

\section*{Acknowledgments}
E.V. and S.M. were supported in part by grants NASA-NNX10AD86G,
HST-AR-12158.01. F.D and A.D. were supported in part by the PRIN-INAF 2009 grant 'Formation and Early Evolution of Massive Star Clusters'.

{}

\end{document}